\documentclass[12pt]{report}
\usepackage[dvips]{graphicx}
\newcommand{\sptwo}{1.4}

\newcommand{\doublespace}{\edef\baselinestretch{\sptwo}\Large\normalsize}

\begin{document}
\doublespace
\begin{center}
{\bf Weizs\"{a}cker energy of unitary Fermi gas in a harmonic trap
}\\
\renewcommand\thefootnote{\fnsymbol{footnote}}
{
Alexander L. Zubarev\footnote{ e-mail: zubareva$@$physics.purdue.edu}}\\
Department of Physics, Purdue University\\
West Lafayette, Indiana  47907\\
\end{center}

\begin{quote}
The universal method of construction of the rigorous lower bounds  to the
Weizs\"{a}cker energy is presented.  We study  a few-fermion system
at the unitarity.  Upper and lower bounds to the density functional theory
 (DFT) ground state energy within the local density approximation (LDA) are
given. The
rigorous lower bounds to the accuracy of the method are  derived.
\end{quote}
\vspace{5mm}

\pagebreak

There has been a lot of interest in systems of fermions at the unitarity [1-3]
(when the scattering length diverges, the Bertsch 
many-body
problem [4]). While Refs.[5-8]
 consider
 homogeneous systems, Refs.[9-13] present {\it ab initio} calculations of the
 properties of trapped fermionic atoms.


The modern DFT is based on the Kohn-Sham approach [14], where the
noninteracting kinetic energy, $T$, is calculated in terms of the Kohn-Sham
orbitals, although Ref.[15] proved the basic existence of functional $T(\rho)$,
where $\rho$ is the density, $\int \rho(\vec{r}) d^3 r=N$, and $N$ is the
 particle number. The accurate density-functional approximation
to the kinetic energy would reduce dramatically complexity of the DFT
calculations (here we note the superfluid extension of the DFT given in
Refs.[16-18]). For applications of the DFT to the nuclear structure physics see web site, 
constructed for the universal nuclear energy density functional
 (UNEDF) collaboration, http://unedf.org.

The kinetic energy functional cam be written as
$$
T[\rho]=\frac{\hbar^2}{2 m}\int \tau(\rho(\vec{r}))d^3 r,
\eqno{(1)}
$$
where 
the semiclassical expansion for the kinetic energy  is [19-23]
$$
T[\rho]=\frac{\hbar^2}{2 m}\int (\tau_{TF}(\rho)+\tau_2(\rho)+\tau_4(\rho)+...)
d^3r,
\eqno{(2)}
$$
where

$$
\tau_{TF}(\rho)=\frac{3}{5} (3 \pi^2)^{2/3} \rho^{5/3},
\eqno{(3)}
$$
$$
\tau_2(\rho)=\frac{1}{36}\frac{(\vec{\nabla} \rho)^2}{\rho},
\eqno{(4)}
$$
$$
\tau_4(\rho)=\frac{1}{6480} (3 \pi^2)^{-2/3} \rho^{1/3} [8 (\frac{\vec{\nabla} \rho}{\rho})^4-27 (\frac{\vec{\nabla} \rho}{\rho})^2 \frac{\triangle \rho}{\rho}+24 (\frac{\triangle \rho}{\rho})^2].
\eqno{(5)}
$$
The Weizs\"{a}cker energy $T_W[\rho]$ [24]
$$
T_W[\rho]=\frac{\hbar^2}{8 m}\int \frac{(\vec{\nabla} 
\rho(\vec{r}))^2}{\rho(\vec{r})}d^3r=\frac{\hbar^2}{2 m}\int 
[\vec{\nabla}\rho^{1/2}(\vec{r})]^2 d^3r,
\eqno{(6)}
$$ 
is an important component of the DFT kinetic energy. Indeed, $T_W$ is considered exact in the
 limit of rapidly varying  density $\rho$  [25,26] and nine times large than
 the second term (the Kirgnitz correction [19]) of the semiclassical expansion
(2), which have to be considered as an asymptotic expansion [23]. 

We present in this letter the universal method to construct the rigorous lower bounds to the $T_W$ which allows us to study a few-fermion system at unitarity.

We rewrite Eq.(6) as
$$
T_W[\rho]=\int[ 
\frac{\hbar^2}{2 m}
[\vec{\nabla}\rho^{1/2}(\vec{r})]^2 +(V(\vec{r},\lambda_1, \lambda_2,...\lambda_M)- V(\vec{r},\lambda_1, \lambda_2,...\lambda_
M))]\rho(\vec{r}) d^3r,
\eqno{(7)}
$$
where $V(\vec{r},\lambda_1, \lambda_2,...\lambda_
M)$ is an one-particle potential and $\lambda$ are parameters.

Introducing an auxiliary Hamiltonian $H_{aux}$ as
$$
H_{aux}= - \frac{\hbar^2}{2 m} \triangle + V(\vec{r},\lambda_1, \lambda_2,...
\lambda_
M),
\eqno{(8)}
$$
and using
$$
<\psi|H_{aux}|\psi> \geq E_G <\psi|\psi>,
\eqno{(9)}
$$
we get
$$
T_W[\rho]\geq N E_G(\lambda_1, \lambda_2,...\lambda_M)
-\int V(\vec{r},\lambda_1, \lambda_2,...\lambda_M)\rho(\vec{r}) d^3r,
\eqno{(10)}
$$
where $E_G$ is the ground state energy of the auxiliary Hamiltonian (8).
Therefore, a set of optimal values of parameters $\lambda_i$ which yield an 
optimal value for a lower bound to the $T_W[\rho]$ is given by
$$
T_W^{lower}[\rho]=\max_{\lambda_i} [N E_G(\lambda_1, \lambda_2,...\lambda_M)
-\int V(\vec{r},\lambda_1, \lambda_2,...\lambda_M)\rho(\vec{r}) d^3r]
\eqno{(11)}
$$

Now let us collect some lower bounds to the Weizs\"{a}cker energy $T_W[\rho]$:

(i) $V(\vec{r},\lambda)=m \lambda^2 r^2/2$, $E_G(\lambda)=(3/2) \hbar \lambda$,
$\lambda \geq 0,$ 
$$
T_W[\rho]\geq \frac{9}{8} \frac{\hbar^2 N^2}{m <r^2>}.
\eqno{(12)}
$$

(ii) $V(\vec{r},\lambda)=-\lambda/r$, $E_G(\lambda)= \lambda^2 m/(2 \hbar^2)$, 
$\lambda \geq 0,$
$$
T_W[\rho]\geq \frac{\hbar^2}{2 m N} <r^{-1}>^2.
\eqno{(13)}
$$

(iii) $V(\vec{r},\lambda_1, \lambda_2,\lambda_3)=(m/2)(\lambda_1^2 x^2+
\lambda_2^2 y^2+ \lambda_3^2 z^2)$, $E_G(\lambda_1, \lambda_2,\lambda_3)=
(\hbar/2)(\lambda_1+\lambda_2+\lambda_3)$, $\lambda_i \geq 0$,
$$
T_W[\rho]\geq \frac{\hbar^2 N^2}{8 m} (\frac{1}{<x^2>}+\frac{1}{<y^2>}+
\frac{1}{<z^2>}).
\eqno{(14)}
$$

(iv) $V(\vec{r},\lambda)=m \lambda^2 r^2/2+\beta(\beta+1)\hbar^2/(2 m r^2)$,
$E_G=\hbar \lambda(3/2+\beta),$ $\beta \geq 0$, $\lambda\geq 0$,
$$
T_W[\rho]\geq \frac {\hbar^2}{2 m}<r^{-2}> \frac{3 N^2+<r^2><r^{-2}>}
{4(<r^2><r^{-2}>-N^2)}.
\eqno{(15)}
$$

(v) $V(\vec{r},\lambda)=-\lambda/r+\beta(\beta+1)\hbar^2/(2 m r^2)$,
$E_G=-\lambda^2 m/(2 \hbar^2 (1+\beta)^2),$ $\beta \geq 0$, $\lambda\geq 0$,
$$
T_W[\rho]\geq \frac {\hbar^2}{8 m} <r^{-2}> \frac{1}
{1-\frac{<r^{-1}>^2}{N<r^{-2}>})},
\eqno{(16)}
$$
where $<r^{\alpha}>=\int\rho(\vec{r}) r^{\alpha} d^3 r$.

The bounds (12), (13), (15) and (16) where previously proved in Refs. [27,28], while the bound (14) is new.
 
In Refs.[29,30] and later in Refs.[31-42] the dynamics of strongly interacting  dilute Fermi gases
 (dilute in the
sense that
 the range of interatomic potential is small
compared with inter-particle spacing) consisting  of a 50-50 mixture of two
 different
 states and
confined in a harmonic trap
$V_{ext}(\vec{r})=(m/2)(\omega_{\perp}^2 (x^2+y^
2)+
\omega_z^2 z^2)$
 is investigated in the single equation approach to the time-dependent
density-functional theory. 

  Let us come back to the variational
formulation of the Kohn-Sham time-dependent theory
$$
\delta \int dt<\psi|i\hbar \partial_t-H|\psi>=0,
\eqno{17)}
$$
where $|\psi>$ is a product of two Slater determinants, one for each internal
 state built up by the Kohn-Sham orbitals $\psi_i$,  and $H=T+U$ is the LDA
Hamiltonian.

Using two approximations

(i) local transform $\psi_i\approx \phi_i \exp (i\hbar \chi/m)$,
where $\psi_i$ and $\chi$ are real functions,

\noindent
and

(ii) $<\phi|T|\phi>\approx \frac{\hbar^2}{2 m}\int \tau_{TF}(\rho)d^3 r+
T_W[\rho],$

where $|\phi>$ is the product of two Slater determinants built on $\phi_i$ 
alone,
we can derive 
the DFT equation of Ref.[29]
$$
i\hbar \frac{\partial \Psi}{\partial t}=-\frac{ \hbar^2}{2 m} \nabla^2 \Psi
+V_{ext} \Psi+V_{xc}\Psi,
\eqno{(18)}
$$

where $V_{xc}(\vec{r},t)=[\frac{\partial \rho \epsilon(
\rho)}{\partial \rho}]_{\rho=\rho(\vec{r},t)}$, $\epsilon(\rho)$ is the 
ground-state energy per particle of the homogeneous system and
$
\rho(\vec{r},t)=\mid \Psi(\vec{r},t)\mid^2.
$
In the case of large but finite number of atoms $N$,
 at small distances the ratio  $\mid \nabla \rho\mid/\rho^{4/3}$ is small and
 both
 the Kirzhnitz correction and the  Weizs\"{a}cker correction are negligible.
On the contrary, 
the Weizs\"{a}cker correction 
is expected to determine the asymptotic
 behavior of the density at large distances. 
As for the case of relatively  small number of atoms, we expect that  the
 Kirzhnitz 
correction would be a reasonable approximation to the kinetic energy
$$
T[\rho]\approx T_{TF}[\rho]+\frac{1}{9} T_W[\rho]
d^3r,
\eqno{(19)}
$$
where $T_{TF}[\rho]=\frac{\hbar^2}{2 m}\int (\tau_{TF}(\rho) d^3r$.

Recently, Ref.[41] has considered the following nonlinear equation
 $$
i\hbar \frac{\partial \Psi_s}{\partial t}=-\frac{ \hbar^2}{4 m} \nabla^2 \Psi
+2 V_{ext} \Psi_s+2 V_{xc}\Psi_s,
\eqno{(20)}
$$
with $|\Psi_s|^2=\rho/2$.

For the stationary case Eq.(20)  reduces to
$$
-\frac{ \hbar^2}{8 m} \nabla^2 \Psi
+V_{ext} \Psi+V_{xc}\Psi=\mu \Psi
\eqno{(21)}
$$
which corresponds to the following approximation of the kinetic energy
$$
T[\rho]\approx T_{TF}[\rho]+\frac{1}{\kappa} T_W[\rho]
\eqno{(22)}
$$
with $\kappa=4$.

For the remainder of this letter we will test the accuracy of the approximation
(22) with $\kappa=9$ and $\kappa=4$ for few-fermion systems at unitarity in a spherical harmonic trap
$$
V_{ext}(\vec{r})=\frac{m \omega^2 r^2}{2}.
\eqno{(23)}
$$

In the approximation (22), the ground state energy is given by the minimum of
 the energy functional
$$
J[\Psi]=\frac{1}{\kappa}T_W[\rho]+\int V_{ext} \rho d^3r+\int \epsilon(\rho) 
\rho d^3r,
\eqno{(24)}
$$
where $\epsilon(\rho)=(1+\beta) 3 \hbar^2 k_F^2/(10 m)$, 
$k_F=(3 \pi^2 \rho)^{1/3}$, $\rho=|\Psi|^2$, and the universal parameter 
$\beta$ is estimated to be $\beta=-0.56$ [7].

Introducing an auxiliary Hamiltonian
$$
\tilde{H}=3 \frac{\hbar \omega}{2} \sqrt{\frac{\lambda}{\kappa}}+
\frac{m \omega^2}{2} (1-\lambda) r^2,
\eqno{(25)}
$$
we can rewrite Eq.(24) as
$$
J[\Psi]=\frac{1}{\kappa}T_W[\rho]+\int (V_{ext}-  \tilde{H}) \rho d^3r 
+\int (\tilde{H}+\epsilon(\rho))\rho d^3r.
$$
Omission of $(1/\kappa)T_W[\rho]+\int (V_{ext}-  \tilde{H}) \rho d^3r$
yields our approximation for the ground state energy
$$
E\approx \frac{3}{2} N \hbar \omega \sqrt{\frac{\lambda}{\kappa}}+
E_{TF}(\sqrt{1-\lambda} \omega),
\eqno{(26)}
$$
where $E_{TF}$ is the Thomas-Fermi energy which is given by
$E_{TF}(\omega)=\sqrt{1+\beta} \hbar \omega (3 N)^{4/3}/4$

Projecting $|\Psi>$on the complete basis states $|n>$, obtained from 
$h|n>=e_n |n>$, where $h=-(\hbar^2/(2 \kappa m))\triangle+m \omega^2 
\lambda r^2/2$, we get
$$
<\Psi|h|\Psi>=\sum_n e_n<\Psi|n><n|\Psi> \geq\frac{3}{2}N \hbar \omega
\sqrt{\frac{\lambda}{\kappa}}.
$$
Therefore, we conclude that our approximation for energy, given by Eq.(26),
 is a lower bound to the ground sate energy, Eq.(24). The optimal value of 
parameter $\lambda$ which maximizes the energy, Eq.(26), will yield an optimal
 value of the lower bounds to the ground state energy given by
$$
\frac{E}{\hbar \omega} \geq \frac{E^{(-)}}{\hbar \omega}=\frac{3}{2}
\frac{N}{\sqrt{\kappa}}\sqrt{1+\frac{(3 N)^{2/3} (1+\beta) \kappa}{4}}.
\eqno{(27)}
$$

For large $N$, the finite $N$ correction can be written as
$$
\frac {E^{(-)}}{\hbar \omega}=\frac{(3 N)^{4/3} \sqrt{1+\beta}}{4}+
\frac{(3 N)^{2/3}}{2 \kappa \sqrt{1+\beta}}+...
\eqno{(28)}
$$
Bhaduri, Murthy and Brack   [43] have recently presented a semiclassical 
approximation, assuming the particles obey the Haldane-Wu fractional 
exclusion statistics at unitarity
$$
 \frac {E^{BMB}}{\hbar \omega}=g^{1/3} \frac{(3 N)^{4/3}}{4}+g^{-1/3}
\frac{(3 N)^{2/3}}{8}+...,
\eqno{(29)}
$$
 where the statistical parameter $g$ is related to the universal parameter 
$\beta$ by the relation $g=(1+\beta)^{3/2}$. We note here that for the case of
 $\kappa=4$ [41] our lower bound, Eq.(28), agrees with the expression of 
Ref.[43].

To calculate upper bounds, $E^{(+)}$, we employing Fetter's trial functions [44]
$$
\rho^{1/2}(\vec{r})=c[1-(1-q)(\gamma r)^2]^{\frac{1}{1-q}},
\eqno{(30)}
$$
where $\gamma$, $q$ are the variational parameters and $c$ is the normalization constant, to minimize the functional $J$, Eq.(24). 

From Table I, we can see that $E^{(-)}>E^{MC}$ for $\kappa=4$ and 
$2\leq N/2\leq 14$. Therefore
$$
\frac{|E(\kappa=4)-E^{MC}|}{E^{MC}}> \triangle(\kappa=4),
$$
where $\triangle(\kappa=4)=(E^{(-)}(\kappa=4)-E^{MC})/E^{MC}$ and
$E(\kappa)$ is the exact ground state solution of Eq.(24).
For $\kappa=9$ and $3\leq N/2\leq 14$, $E^{(+)}\leq E^{MC}$ (see Table I),
that is why
$$
\frac{|E^{MC}-E(\kappa=9)|}{E^{MC}}> \triangle(\kappa=9),
$$
where
$\triangle(\kappa=9)=(E^{MC}-E^{(+)}(\kappa=9))/E^{MC}.$ 

We can, therefore, state that $\triangle(\kappa)$ is the rigorous lower  
bound to the accuracy of the approximation (22) with $\kappa=4$, 
$2\leq N/2\leq 14$
and $\kappa=9$, $3\leq N/2\leq 14$. As for the lower $E^{(-)}$ and the
upper  $E^{(+)}$  bounds, they provide the actual solution of Eq.(24),
$(E^{(+)}+E^{(-)}/2$ within $\pm \delta$ accuracy, with $\delta<1\%$ for both 
$\kappa=9$ and $\kappa=4$ and for $2\leq N/2\leq 15$.

The predictions of Eq.(24) for the ground state energy 
and results of Ref.[10] are shown in Fig.1. A very good agreement between
 the Kirgnitz approximation, $\kappa=9$, and the {\it ab initio} calculations
 of Ref.[10] 
can be seen for $2\leq N/2 \leq 4$. However, for $12 \leq N/2 \leq 15$ the
$\kappa=4$ approximation gives better results than the $\kappa=9$ 
approximation. 

Finally, we note that
the kinetic energy functionals assumed here are not unique. In
our future work we will consider other possible forms.

In summary, we have constructed the rigorous lower bounds to the Weizs\"{a}cker energy. As example of application, we have studied few-fermion systems at unitarity consisting of 50-50 mixture of two different states and confined in a spherical harmonic trap. The rigorous lower bounds to the accuracy of the method are derived. We have tested the kinetic energy functionals by comparisons with
{\it ab initio} calculations and have found that while the second order 
gradient expansion is a very accurate for relatively  small $N$, Fig.1 
indicates 
that the $\kappa=4$ approximation provides significantly better results for
 larger $N$.

\noindent
{\bf Acknowledgments}

I thank N.J. Giordano and W.L. Fornes for providing the opportunity to finish 
this work.

\pagebreak

Table I. The energies $E^{(-)}$, $E^{(+)}$ and the energy  calculated within
 the 
fixed-node diffusion Monte Carlo method, $E^{MC}$ [10], all in units of 
$\hbar \omega$
for $N\leq 30$
[see the text for further details].

\vspace{8pt}

\noindent
\begin{tabular}{llllllll}
\hline\hline
$N/2$
&$E^{(-)}(\kappa=4)$
&$E^{(+)}(\kappa=4)$
&$E^{MC}$ 
&$E^{(-)}(\kappa=9)$
&$E^{(+)}(\kappa=9)$
&$\triangle(\kappa=4)$
&$\triangle(\kappa=9)$ \\ \hline
2
&5.455
&5.561
&5.05
&4.976
&5.062
&8.0\%
&-  \\ \hline
3
&9.025
&9.198
&8.64
&8.378
&8.513
&4.5\%
&1.5\% \\ \hline
4
&12.954
&13.196
&12.58
&12.157
&12.341
&3.0\%
&1.9\% \\ \hline
5
&17.182
&17.495
&16.81
&16.247
&16.479
&2.2\%
&2.0\% \\ \hline
6
&21.670
&22.055
&21.28
&20.605
&20.885
&1.8\%
&1.9\%  \\ \hline
7
&26.390
&26.846
&25.92
&25.202
&25.530
&1.8\%
&1.5\%  \\ \hline
8
&31.318
&31.848
&30.88
&30.014
&30.389
&1.4\%
&1.6\%  \\ \hline
9
&36.440
&37.043
&35.97
&35.024
&35.445
&1.3\%
&1.5\%  \\ \hline
10
&41.741
&42.416
&41.30
&40.216
&40.684
&1.1\%
&1.5\%  \\ \hline
11
&47.208
&47.956
&46.89
&45.578
&46.093
&0.7\%
&1.7\%  \\ \hline
12
&52.833
&53.653
&52.62
&51.100
&51.662
&0.4\%
&1.8\%  \\ \hline
13
&58.605
&59.499
&58.55
&56.775
&57.381
&0.1\%
&2.0\%  \\ \hline
14
&64.519
&65.485
&64.39
&62.592
&63.244
&0.2\%
&1.8\%  \\ \hline
15
&70.567
&71.606
&70.93
&68.546
&69.243
&-
&2.4\% 
\\ \hline\hline
\end{tabular}

\pagebreak
\begin{figure}[ht]
\includegraphics{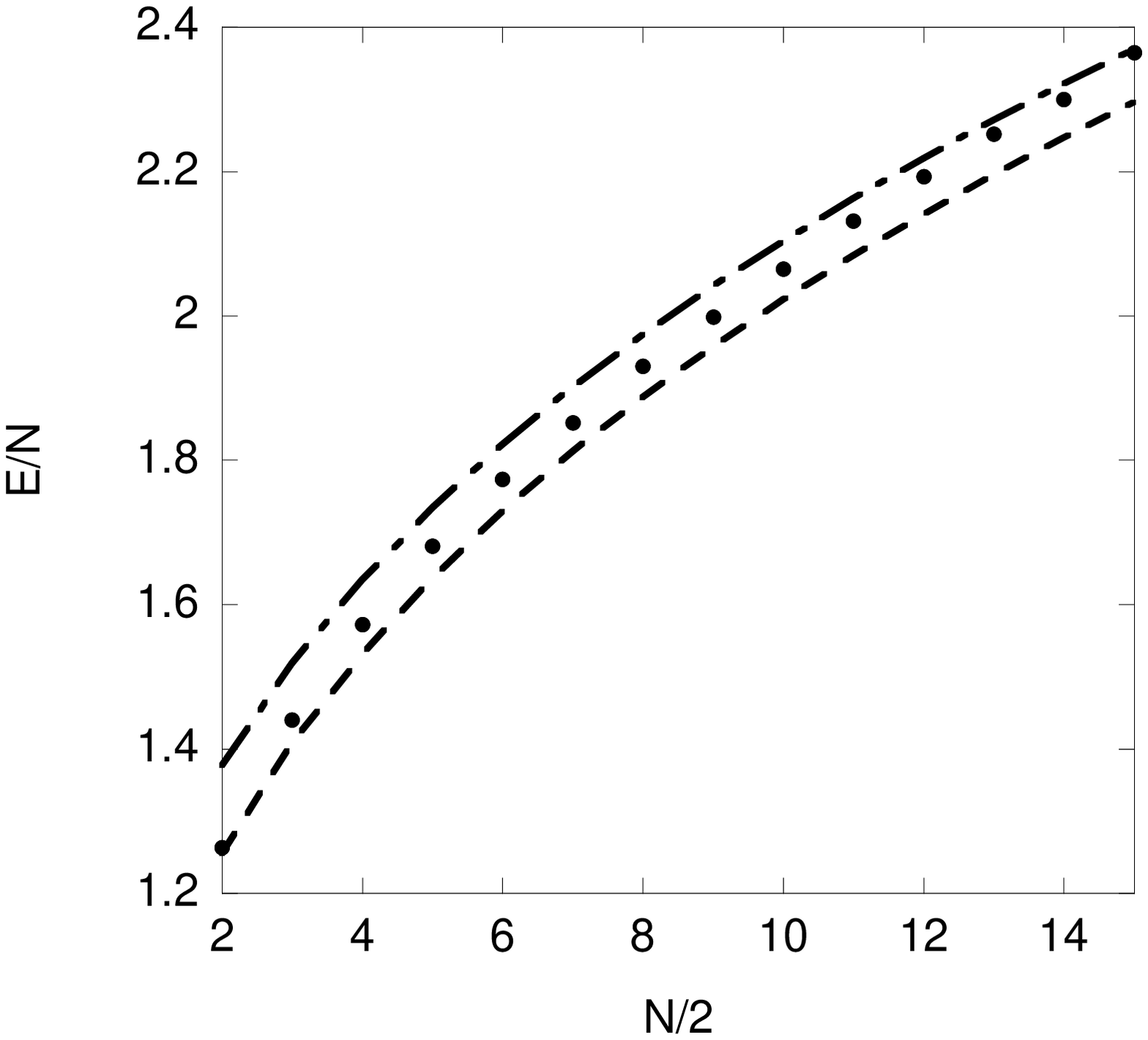}
\end{figure}
Fig.1. Ground state energy per particle of few-fermion systems at unitarity in a spherical harmonic trap in units of $\hbar \omega$ as a function of number 
of atoms $N$. The dashed and the dotted-dashed lines represent $(E^{(-)} 
+(E^{(+)})/2$ for $\kappa=9$ and $\kappa=4$, respectively. The circular dots indicate results of Ref.[10].

\pagebreak

\noindent
{\bf References}

\vspace{8pt}

\noindent
[1] O'Hara K M,  Hemmer S L,  Gehm M E,  S R Granade S R, and
Thomas J E 2002 {\it Science} {\bf 298},2179

\noindent
[2] Bartenstein M,  Altmeyer A, Riedl S, Jochim S, Chin C,
Denschlag J H, and Grimm R  2004 {\it Phys. Rev. Lett.} {\bf 92}, 120401

\noindent
[3] Bourdel T, Khaykovich L, Cubizolles J, Zhang J,
Chevy F, Teichmann M, Tarruell L,
Kokkelmans S J J M F, and Salomon C 2004 {\it Phys. Rev. Lett.} {\bf 93}, 050401

\noindent
[4] Bishop R F 2001 {\it  Int. J. Mod. Phys.} {\bf 15}, iii

\noindent
[5] Astrakharchik G E, Boronat J, Casulleras J,
 and Giorgini S  2004 {\it Phys. Rev. Lett.} {\bf 93}, 200404

\noindent
[6] Chang S Y and Pandharipande V R 2005 {\it Phys. Rev. Lett.} {\bf 95},
080402

\noindent
[7] Carlson J, Chang S-Y, Pandharipande V R, and
Schmidt K E 2003 {\it Phys. Rev. Lett.} {\bf 91}, 050401

\noindent
[8] Carlson J  and  Reddy S 2005 {\it Phys. Rev. Lett.} {\bf 95}, 060401

\noindent
[9] Chang S Y, and Bertsch G F 2007 {\it Phys. Rev.} A{\bf 76}, 021603(R)

\noindent
[10] Blume D,  von Stecher J, and Greene C H 2007 {\it Phys. Rev. Lett.}
{\bf 99}, 233201

\noindent
[11] von Stecher J, Greene C H, and  Blume D 2007 {\it Phys. Rev.} A{\bf 76},
053613

\noindent
[12] Blume D  2008 {\it Phys. Rev.} A{\bf 78}, 013635

\noindent
[13] von Stecher J, Greene C H, and  Blume D 2008 {\it Phys. Rev.} A{\bf 77},
043619

\noindent
[14] Kohn W, and Sham L J 1965 {\it Phys. Rev.} {\bf 140}, A1133

\noindent
[15] Hohenberg P, and Kohn W  1964 {\it Phys. Rev.} {\bf 136}, B1133

\noindent
[16] Yu Y, and Bulgac A 2003 {\it Phys. Rev. Lett.} {\bf 90}, 222501

\noindent
[17] Bulgac A  2002 {\it Phys. Rev.} C{\bf 65}, 051305(R)

\noindent
[18] Bulgac A  2007 {\it Phys. Rev.} A{\bf 76}, 040502(R)

\noindent
[19] Kirgnitz D A 1967 {\it Field Theoretical Methods in Many Body Systems} 
(London: Pergamon Press)

\noindent
[20] Brack M, and Bhaduri R K 1997 {\it Semiclassical Physics} (Reading MA:
 Addison-Wesley)

\noindent
[21] Brack M, Jennings B K, and Chu P H 1976 {\it Phys. Lett.} B{\bf 65}, 1

\noindent
[22] Hodges C H 1973 {\it Can. J. Phys} {\bf 51}, 1428

\noindent
[23] Guet C, and Brack M  1980 {\it Z. Phys.} A{\bf 297}, 247 

\noindent
[24] von Weizs\"{a}cker C F 1935 {\it Z. Phys.} {\bf 96}, 431

\noindent
[25] Parr R G, and Yang W 1989 {\it Density-Functional Theory of Atoms and 
Molecules} (New York: Oxford University Press)

\noindent
[26] Dreizler R M, and Gross E K U 1990 {\it Density Functional Theory: 
An Approach to the Quantum Many-Body Problem} (Berlin: Springer-Verlag

\noindent
[27] Gadre S R and Pathak R K  1982 {\it Phys. Rev.} A{\bf 25}, 668

\noindent
[28] Romera E and Dehesa J S  1994 {\it Phys. Rev.} A{\bf 50}, 256

\noindent
[29] Kim Y E   and Zubarev A L  2004 {\it Phys. Rev.} A{\bf 70}, 033612

\noindent
[30] Kim Y E   and Zubarev A L  2004 {\it Phys. Lett.} A{\bf 327}, 397

\noindent
[31] Kim Y E   and Zubarev A L  2005 {\it Phys. Rev.} A{\bf 72}, 011603(R)

\noindent
[32] Kim Y E   and Zubarev A L  2005 {\it J. Phys. B: At. Mol. Opt. Phys.} 
{\bf 38}, L243

\noindent
[33]  Manini N and Salasnich L 2005 {\it Phys. Rev.} A{\bf 71}, 033625

\noindent
[34] Ghosh  T K and  Machida K 2006 {\it Phys. Rev.} A{\bf 73}, 013613

\noindent
[35] Diana G, Manini N, and Salasnich L 2006 {\it Phys. Rev.} A{\bf 73},
065601 

\noindent
[36] Yin J  and Ma Y-L 2006 {\it Phys. Rev.} A{\bf 74}, 013609

\noindent
[37] Salasnich L and Manini N 2007 {\it Laser Phys.} {\bf 17}, 169

\noindent
[38] Zhou Y and Huang G 2007 {\it Phys. Rev.} A{\bf 75}, 023611

\noindent
[39] Ma Y-L  and Huang G 2007 {\it Phys. Rev.} A{\bf 75}, 063629

\noindent
[40] Wen W and  Huang G 2007 {\it Phys. Lett.} A{\bf 362}, 331

\noindent
[41] Wen W,  Zhou Y and Huang G 2008 {\it Phys. Rev.} A{\bf 77}, 033623

\noindent
[42] Adhikari S K  2008 {\it Phys. Rev.} A{\bf 77}, 045602

\noindent
[43] Bhaduri R K, Murthy M V N and Brack M 2008 
{\it J. Phys. B: At. Mol. Opt. Phys.} {\bf 41}, 115301

\noindent
[44] Fetter A.L. 1997 {\it J. Low. Temp. Phys.} {\bf 106}, 643

\end{document}